\documentclass[twocolumn]{aastex631}

\newcommand\ha{H$\alpha$}
\newcommand\sv{SHARK-VIS}
\newcommand\mj{M$_{J}$}
\newcommand\mic{$\mu$m}
\newcommand\msun{$M_\odot$}
\newcommand\lsun{$L_\odot$}

\shorttitle{A candidate binary companion for CQ Tau}
\shortauthors{Antoniucci et al.}

\usepackage{soul}

\begin{document}

\title{A candidate binary companion in the CQ Tau transitional disk}


\author[0000-0002-0666-3847]{Simone Antoniucci}
\affiliation{INAF - Osservatorio Astronomico di Roma, Via Frascati 33, 00078, Monte Porzio Catone, Italy}
\affiliation{ADONI - Italian National Institute for Adaptive Optics}

 \author[0000-0001-9539-2112]{Gianluca Li Causi}
 \affiliation{INAF - Osservatorio Astronomico di Roma, Via Frascati 33, 00078, Monte Porzio Catone, Italy}
 \affiliation{ADONI - Italian National Institute for Adaptive Optics}

 \author[0000-0003-4679-4072]{Daniele Fasano}
 \affiliation{INAF - Osservatorio Astronomico di Roma, Via Frascati 33, 00078, Monte Porzio Catone, Italy}

 \author[0009-0007-1759-1985]{Maciej W. Bajor}
 \affiliation{Dipartimento di Fisica, Università di Roma Tor Vergata, Via della Ricerca Scientifica 1, 00133 Roma, Italy}
 \affiliation{Dipartimento di Fisica, Sapienza Università di Roma, Piazzale Aldo Moro 5, 00185 Roma, Italy}
 \affiliation{INAF - Osservatorio Astronomico di Roma, Via Frascati 33, 00078, Monte Porzio Catone, Italy}

 \author[0000-0002-0983-8040]{Fernando Pedichini}
 \affiliation{INAF - Osservatorio Astronomico di Roma, Via Frascati 33, 00078, Monte Porzio Catone, Italy}
 \affiliation{ADONI - Italian National Institute for Adaptive Optics}
 
 \author[0009-0000-5623-5237]{Simone Filomeno}
 \affiliation{Dipartimento di Fisica, Università di Roma Tor Vergata, Via della Ricerca Scientifica 1, 00133 Roma, Italy}
 \affiliation{Dipartimento di Fisica, Sapienza Università di Roma, Piazzale Aldo Moro 5, 00185 Roma, Italy}
 \affiliation{INAF - Osservatorio Astronomico di Roma, Via Frascati 33, 00078, Monte Porzio Catone, Italy}

 \author[0000-0001-8467-1933]{Dino Mesa}
 \affiliation{INAF - Osservatorio Astronomico di Padova, Vicolo dell’Osservatorio 5, 35122, Padova, Italy}

 \author[0000-0001-5378-7749]{Enrico Ragusa}
 \affiliation{Dipartimento di Fisica, Università degli Studi di Milano, Via Celoria 16, 20133 Milano, MI, Italy}
 \affiliation{Dipartimento di Matematica, Università degli Studi di Milano, Via Saldini 50, 20133 Milano, Italy}
 
 \author[0000-0002-1892-2180]{Katia Biazzo}
 \affiliation{INAF - Osservatorio Astronomico di Roma, Via Frascati 33, 00078, Monte Porzio Catone, Italy}

 \author[0000-0002-7035-8513]{Teresa Giannini}
 \affiliation{INAF - Osservatorio Astronomico di Roma, Via Frascati 33, 00078, Monte Porzio Catone, Italy}

 \author[0000-0002-9190-0113]{Brunella Nisini}
 \affiliation{INAF - Osservatorio Astronomico di Roma, Via Frascati 33, 00078, Monte Porzio Catone, Italy}

 \author[0000-0002-6796-1100]{Francesco D'Alessio}
 \affiliation{INAF - Osservatorio Astronomico di Roma, Via Frascati 33, 00078, Monte Porzio Catone, Italy}

 \author[0000-0002-6894-1267]{Ester Marini}
 \affiliation{INAF - Osservatorio Astronomico di Roma, Via Frascati 33, 00078, Monte Porzio Catone, Italy}

 \author[0000-0002-2264-8698]{Roberto Piazzesi}
 \affiliation{INAF - Osservatorio Astronomico di Roma, Via Frascati 33, 00078, Monte Porzio Catone, Italy}
 \affiliation{ADONI - Italian National Institute for Adaptive Optics}

 \author[0000-0003-1033-1340]{Vincenzo Testa}
 \affiliation{INAF - Osservatorio Astronomico di Roma, Via Frascati 33, 00078, Monte Porzio Catone, Italy}
 \affiliation{ADONI - Italian National Institute for Adaptive Optics}

 \author[0009-0003-7811-1683]{Piero Vaccari}
 \affiliation{INAF - Osservatorio Astronomico di Roma, Via Frascati 33, 00078, Monte Porzio Catone, Italy}
 \affiliation{ADONI - Italian National Institute for Adaptive Optics}

 \author[0000-0002-5840-8362]{Jacopo Farinato}
 \affiliation{INAF - Osservatorio Astronomico di Padova, Vicolo dell’Osservatorio 5, 35122, Padova, Italy}
 \affiliation{ADONI - Italian National Institute for Adaptive Optics}

 \author[0000-0002-8822-6834]{Luca Marafatto}
 \affiliation{INAF - Osservatorio Astronomico di Padova, Vicolo dell’Osservatorio 5, 35122, Padova, Italy}

 \author[0000-0002-4977-0378]{Elena Carolo}
 \affiliation{INAF - Osservatorio Astronomico di Padova, Vicolo dell’Osservatorio 5, 35122, Padova, Italy}

 \author[0009-0005-9862-2549]{Domenico Barbato}
 \affiliation{INAF - Osservatorio Astronomico di Padova, Vicolo dell’Osservatorio 5, 35122, Padova, Italy}

 \author[0000-0003-1949-6010]{Simone Di Filippo}
 \affiliation{INAF - Osservatorio Astronomico di Padova, Vicolo dell’Osservatorio 5, 35122, Padova, Italy}

 \author[0000-0001-7819-9003]{Cecilia Lazzoni}
 \affiliation{INAF - Osservatorio Astronomico di Padova, Vicolo dell’Osservatorio 5, Padova, Italy, 35122}

 \author{Lorenzo Merli}
 \affiliation{INAF - Osservatorio Astronomico di Padova, Vicolo dell’Osservatorio 5, 35122, Padova, Italy}

 \author[0009-0002-2503-3186]{Alessandro Lorenzetto}
 \affiliation{INAF - Osservatorio Astronomico di Padova, Vicolo dell’Osservatorio 5, 35122, Padova, Italy}

 \author{Cristina Knapic}
 \affiliation{INAF - Osservatorio Astronomico di Trieste, Via Tiepolo, 11, 
 34131, Trieste, Italy}

 \author{Grant West}
 \affiliation{Steward Observatory and Department of Astronomy, University of Arizona, 933 N Cherry Avenue, Tucson, AZ 85721, USA}
 
 \author[0000-0002-6243-5697]{Enrico Pinna}
 \affiliation{INAF – Osservatorio Astrofisico di Arcetri, Largo E. Fermi 5, 50125 Firenze, Italy}
 \affiliation{ADONI - Italian National Institute for Adaptive Optics}

 \author[0000-0002-3544-1629]{Fabio Rossi}
 \affiliation{INAF – Osservatorio Astrofisico di Arcetri, Largo E. Fermi 5, 50125 Firenze, Italy}
 \affiliation{ADONI - Italian National Institute for Adaptive Optics}

\begin{abstract}

High-contrast observations of pre-main sequence stars are key to investigate planet formation and disk evolution. 
Here, we present observations of the CQ Tau transitional disk, taken with the high-contrast instruments SHARK-VIS and SHARK-NIR of the Large Binocular Telescope (LBT) in \ha\ and $H$-band, respectively, aimed at searching for young giant planets in formation within the disk.
We detect no point-like source associated with forming planets. From the \ha\ and near-infrared contrast curves we derive limits on the accretion emission and mass of possible planetary companions, ruling out objects with $M>5 M_J$ at separations $>$ 70 au. 
An optimized procedure to inspect the central region of the SHARK-VIS images, taken without coronagraph, shows however a candidate stellar companion at very close 
separation (27.4 mas, corresponding to $a$ = 5.0 au on the plane of the outer disk). 
From its contrast, we infer a mass estimate in the range 0.46-0.85 \msun.
Our scattered-light images retrieve several structures of the disk, revealing additional spiral arms compared to previous high-contrast observations, a twisted morphology of the bright NE arm, and confirm the presence of wide shadows cast by an inner misaligned disk. 

The candidate companion might be the object invoked by previous works to explain the disturbed disk kinematics and its spirals. In alternative scenarios, some disk structures might be connected to late matter infall from the ambient medium or to mass loss at constant speed creating spiral arms with a constant pitch, such as those observed in the NW section of the disk. 

\end{abstract}




\section{Introduction}
\label{sec:intro}

In recent years, high-resolution observational campaigns conducted with extreme adaptive optics (AO) systems and interferometers have revealed that protoplanetary disks display a wide range of substructures, such as rings, gaps, cavities, and spiral arms, which strongly suggest that planet formation is
already active in the early stages of disk evolution \citep[e.g.][]{benisty23}.
Although alternative hypotheses exist to explain the occurrence of those structures, such as spirals driven by disk self-gravity or gravitational instability \citep[e.g.][]{dong18, Speedie24}, illumination-induced density waves from inner-disk shadows \citep[e.g.][]{montesinos16, Ziampras25}, or perturbations from stellar flybys \citep[e.g.][]{cuello19}, the presence of forming massive planets remains the most frequently invoked mechanism to clear and sculpt sections of the disks. 
In particular, transitional disks, characterized by large inner cavities depleted of dust, are considered ideal laboratories for studying planet formation processes \citep[e.g.][]{vandermarel23}, as their peculiar morphology is frequently attributed to the dynamical interaction with hidden stellar or sub-stellar companions \citep[e.g.][]{dong15a,bae23,ragusa25}.

While near-infrared AO observations have successfully mapped the scattered light from the surface of many disks, they have so far mostly failed to directly detect planets, because of the intrinsic difficulty in spotting and disentangling the weak planet signal from the features of the disk. To date, the giant planets of the PDS 70 \citep{keppler18,wagner18,haffert19} and WISPIT 2 \citep{vancapelleveen25,close25b,lawlor26} systems remain the only widely accepted protoplanets, with a few other candidates that still need further validation, e.g. AB Aur b \citep{currie22}, HD 169142 b \citep{hammond23}, HD 135344 b \citep{maio25}.

In this context, adding high-contrast imaging in the visible band to standard near-infrared data provides a complementary way to detect forming protoplanets via their bright \ha\ line \citep[e.g.][]{haffert19,close25b}, which is emitted during their gas accretion phase \citep[e.g.][]{aoyama21,marleau22}. Moreover, visible observations naturally offer enhanced resolution, which is key to probe the innermost regions, where the bulk of the giant planets are expected to form \citep[e.g.][]{fulton21,emsenhuber21b}.

In this work, we present new \ha\ and near-infrared continuum observations of the pre-main sequence star CQ Tau, obtained with the novel high-contrast instruments SHARK-VIS \citep{pedichini22,pedichini24} and SHARK-NIR \citep{2015IJAsB..14..365F,2022SPIE12185E..22F,2022SPIE12184E..3VM} of the Large Binocular Telescope (LBT), operating downstream of the SOUL extreme adaptive optics \citep{pinna23}.
The data were collected in the framework of our new SHARK program to search for giant gaseous protoplanets in the Taurus-Auriga star-forming region, leveraging the optimal location of the LBT for AO observations of targets in this northern cloud.

CQ Tau is an intermediate-mass pre-main sequence star \citep[$T_\mathrm{eff}\sim$ 6800 K, $M = 1.5\mathrm{-}1.7$ $M_\odot$, e.g.][and references therein]{alcala21,hammond22} located in the Taurus star-forming complex at a distance of 149.4 pc \citep[][]{gaia21}, with age estimates ranging from 8 to 12 Myr \citep[e.g.][]{hammond22, garufi24}.
The star is surrounded by a massive disk that has been the subject of intense multi-wavelength investigations, revealing an extremely complex architecture.
Using ALMA observations on continuum emission and CO lines, \citet{ubeira-gabellini19} found a large cavity in the millimeter-sized dust distribution with a radius around 52-53 au, and a smaller gas cavity extending to 20–25 au. They first proposed that this large gap could be cleared by an unseen planet with a mass of 6–9 M$_{J}$ located at a semi-major axis of 20 au.

Subsequent scattered light images in the near-infrared \citep{uyama20,hammond22} uncovered prominent spiral arms developing from inside the millimeter dust cavity. These data also revealed illumination asymmetries and broad shadows, which might be connected to the presence of a compact slightly misaligned inner disk identified through interferometric measurements \citep{bohn22,winter25}.

Further analysis of the CO kinematics with ALMA by \citet{woelfer21} and more recently by \citet{benisty26} have highlighted spirals and strong vertical flows in the gas, providing further support for the hypothesis of a massive embedded companion that influences the disk gas dynamics. At the same time, \citet{Pinte25} reported large velocity deviations from Keplerian motion consistent with the presence of a massive companion, above the planetary limit \citep[e.g.][]{calcino23, calcino24}.

CQ Tau also exhibits irregular photometric variability typical of UX Ori stars, usually attributed to transient dust clouds periodically obscuring the line of sight. Recent analyses of historical light curves spanning 125 years \citep{grinin23} have uncovered a 10-year periodic component, which the authors connect to the periodic perturbations in the circumstellar material induced by the orbital motion of a possible close-in stellar companion.

Finally, \citet{vioque26} recently performed a new analysis of Gaia data
and identified a significant proper motion anomaly for CQ Tau, consistent with a close-in brown dwarf or stellar companion. 

Despite all these indirect observational clues, all previous searches using high-contrast imaging in the infrared \citep{uyama20,safonov22,hammond22,garufi24} failed to reveal any stellar or sub-stellar companion to CQ Tau. 
Limits set by \citet{hammond22} rule out giant planets above 4 \mj\ immediately outside the spiral region ($\sim$ 60 au) and above 2-3 \mj\ beyond 100 au.

We point out that the SHARK-VIS images we present here are the first high-contrast observations of CQ Tau obtained in the visible band. As for all the targets of our Taurus-Auriga survey, the SHARK-VIS data are specifically acquired without a coronagraph to take advantage of the enhanced spatial resolution offered by the visible light ($\sim$20 mas at the LBT), so as to probe the innermost regions of the sources for (sub-)stellar or planetary companions. 
 
Our paper is structured as follows: in Sect. \ref{sec:obs} we give a description of the observations and data reduction; in Sect. \ref{sec:results} we present the final images and analyze them to obtain our results; we discuss our findings in Sect. \ref{sec:discussion} in the context of previous observations of CQ Tau; finally, we summarize our conclusions in Sect. \ref{sec:conclusions}.\\

\section{Observations and data reduction}
\label{sec:obs}

\subsection{SHARK-VIS data}
\label{sec:visobs}

We observed CQ Tau with SHARK-VIS, operating on the right telescope of the LBT, on UT 2024 February 17, during a commissioning run of the instrument, using two filters: a narrow-band \ha\ filter (656.3 nm, 1 nm bandwidth) and an adjacent continuum filter (647 nm, 10 nm bandwidth).
The continuum frames were acquired to provide a control dataset and to assess the presence of sources emitting exclusively in \ha.

We used a field stop with a diameter of 2.4 arcsec and acquired a total of 12000 512$\times$512 pixel raw frames in H$\alpha$, interspersed with 1200 raw frames in the continuum, switching every two minutes between the filters\footnote{SHARK-VIS now offers a dedicated dual-band mode with an optimized beam-splitter for simultaneous \ha\ and continuum imaging \citep{pedichini24}, but this configuration was not yet available during the commissioning run in which CQ Tau was observed.}.
Each frame had an integration time (DIT) of 0.5 s, resulting in total exposure times of 1h 40m and 10m for \ha\ and continuum, respectively. 

The observations were performed in pupil-stabilized mode around culmination of the target to maximize field rotation for the application of the Angular Differential Imaging technique (ADI; \citet{marois2006}); the total field rotation during the observation was 115.5 degrees. Atmospheric conditions were relatively stable, with a median seeing of about $0\farcs9$.

The SHARK-VIS frames were acquired with an ANDOR Zyla 4.2 Plus sCMOS science camera, which provided a plate scale of 6.495 mas/pix. Data were reduced with the custom DReAMS (Data Reduction, Analysis, and Modeling for SHARK-VIS) pipeline (Li Causi et al. in prep), developed in-house by our group.

The initial reduction steps, which are aimed at cleaning the frames from detector signatures, included:
i) subtraction of a cleaned average dark image, separately acquired at the end of the observing night;
ii) correction for frame-by-frame variable bias patterns, mainly column-by-column (and to a lesser extent row-by-row) non-static bias differences, measured on each frame as column/row averages of pixels outside the field stop (while all pixels were used for dark frame cleaning);
iii) masking of cosmic rays (CR), identified via a wide-box sigma filter on the maximum-pixel frame stack after subtraction of its radial profile, and refined with a temporal sigma filter applied to the candidate CR pixels;
iv) flat-field correction, using a pixel non-uniform response map acquired with the SHARK-VIS internal calibration unit.

We then performed a Gaussian peak analysis of the cleaned frames to reject frames with poor peak quality. The selected frames were co-aligned via Fourier translation using the measured peak locations. Frames with anomalous peak shapes (e.g., double-peaked or motion-blurred) were also discarded. 

After these cleaning procedures, we performed a final frames selection discarding the worst frames on the basis of their peak FWHM, intensity, and ellipticity, finally retaining 11,365 H$\alpha$ frames (95\% of the raw set) and 1,119 continuum frames (93\%), corresponding to about 1h 35m and 9.3m of effective exposure time, respectively. The quality of the H$\alpha$ average stacked image is good, with a full width at half maximum (FWHM) of 27 mas (4.11 pixels), corresponding to a Strehl ratio of about 13\%, and a background contrast within the AO control radius\footnote{This is the radius that identifies the circular region within which the AO correction is effective, which ultimately depends on the number of modes corrected by the AO system and on the wavelength of observation. Outward of the control radius we start to observe the residual of the original seeing profile; this is apparent especially in the averaged images, which present a bright annulus corresponding to the transition from the AO-corrected region to the uncorrected region. This brighter annulus is typically found between 230 and 300 mas for good quality \sv\ observations at LBT.} of $2.5\times10^{-3}$ at 200 mas.

For both filters, we then applied the ADI with Principal Component Analysis (PCA-ADI; \citet{amara_quanz_2012}, \citet{soummer2012}) to subtract the stellar PSF.

\begin{figure*}[t!]
    \centering
    \includegraphics[width=0.92\linewidth]{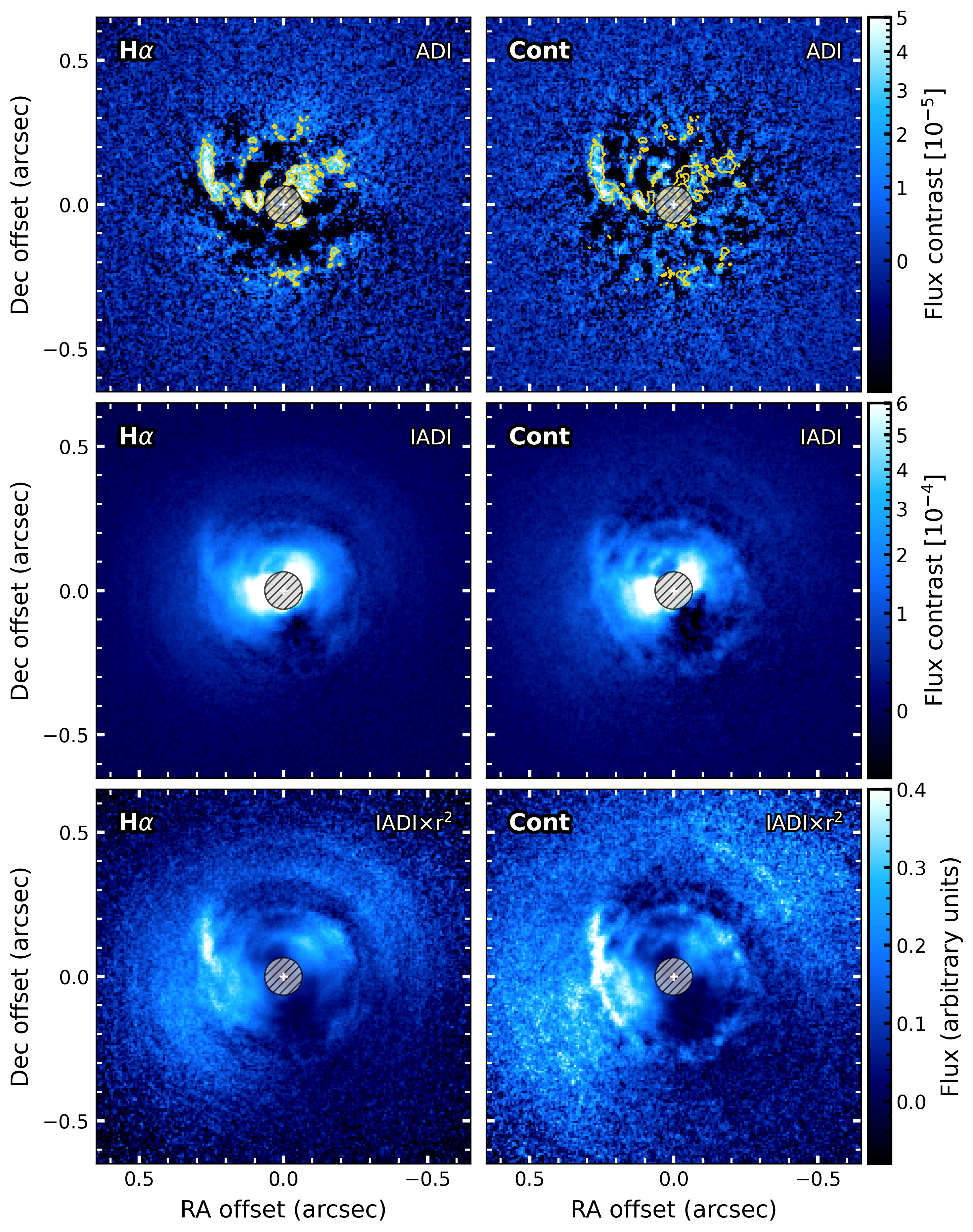}
     \caption{SHARK-VIS images of CQ Tau for \ha-1 nm (left) and continuum 647-10 nm (right). The images are displayed with North on top, East on the left, and a square root color scale with pixel values expressed in contrast. \textit{Top row:} PCA-ADI with 100 components. \ha\ contours at 8 $\sigma$ are overlaid on both \ha\ and continuum panels. \textit{Mid row:} PCA-IADI iterating components from 2 to 75 for \ha\ and from 2 to 45 for continuum (see text for details). \textit{Bottom row:} PCA-IADI multiplied by $r^2$. The central circle marks the area of unreliable signal due to the attenuation caused by the high number of PCA components. 
     }
\label{fig:adi_iadi_r2}
\end{figure*}

We tried different numbers of PCA components to model the PSF for subtracting the central star in the ADI process,
eventually using 100 components for both filters, which results in the best compromise between the compensation of the ADI self-subtraction and the increase of the PCA attenuation in the inner region.
The ADI result for the continuum $647~\rm nm$ is affected by a much larger noise, due to the smaller frame statistics (for this commissioning observation, we only acquired one tenth of the frames in the continuum compared to \ha) not allowing the PCA to finely model the PSF variations.
The ADI images show only a few bright sections of the disk structures, that are affected by non-negligible self-subtraction, as well as a number of condensed residuals (Fig. \ref{fig:adi_iadi_r2}, top row).

To improve the ADI results, we performed a PCA-based Iterative Angular Differential Imaging (PCA-IADI; \cite{pca_iadi}, \cite{Pairet_2021}), in which the structures in the ADI result are iteratively subtracted from the frames before re-launching the ADI process, a method that yields a PSF model substantially unaffected by the circumstellar signal, thus drastically reducing both attenuation and self-subtraction effects.
In our IADI process we considered as model for the object all the positive pixels after each iteration, and used an iteration scheme in which the number of PCA components is incremented by one at each step starting from component 2. After examination of the images, we chose to limit the iterations to 75 components for \ha\ and 45 components for continuum.
This procedure effectively recovered the flux from several structures of the disk (Fig. \ref{fig:adi_iadi_r2}, middle row).

\subsection{SHARK-NIR data}
\label{sec:nirobs}

In parallel with the SHARK-VIS observations we performed simultaneous observations with the twin instrument SHARK-NIR, operating on the left side of the LBT. The observations were carried out using the Gaussian coronagraph (inner working angle 0\farcs12) and the broad-band H filter (BB\_H with a central 
wavelength of 1.6~\mic~and a bandwidth of 0.218~\mic). We obtained 400 science frames with an exposure time of 16.05~s each for a 
total time on target of 6420.24~s ($\sim$107~minutes). The pixel scale was $14.5$ mas/pixel, and the observation was performed in pupil-stabilized mode allowing the rotation
of the field of view to be able to perform ADI. The total field rotation was $112.3^{\circ}$. Before and after the scientific 
frames we also took frames with the stellar PSF outside the coronagraph with the aim of correctly estimating the contrast at different 
separations from the central star. Such frames were taken using an appropriate neutral density filter (in this case a ND3 filter) to
avoid saturation of the detector. \par
The data were reduced by creating and subtracting a master dark frame and dividing by the master flat frame.
To precisely define the position of the stellar PSF behind the coronagraph, we exploited the grid of spots created by 
the internal deformable mirror (DM) of SHARK-NIR. In particular, we used the four brightest spots that are in a cross configuration 
around the stellar PSF, fitting two lines passing through two opposite spots and considering the intersection of such lines as the 
position of the star. Furthermore, we identified and excluded all the bad quality scientific frames, such as
those where the PSF was not adequately masked by the coronagraph because of residual AO jitter. After this procedure, we remained with 
353 science frames for a total exposure time of 5665.8~s ($\sim$94~minutes). 

\begin{figure*}
    \centering
    \includegraphics[width=0.92\linewidth]{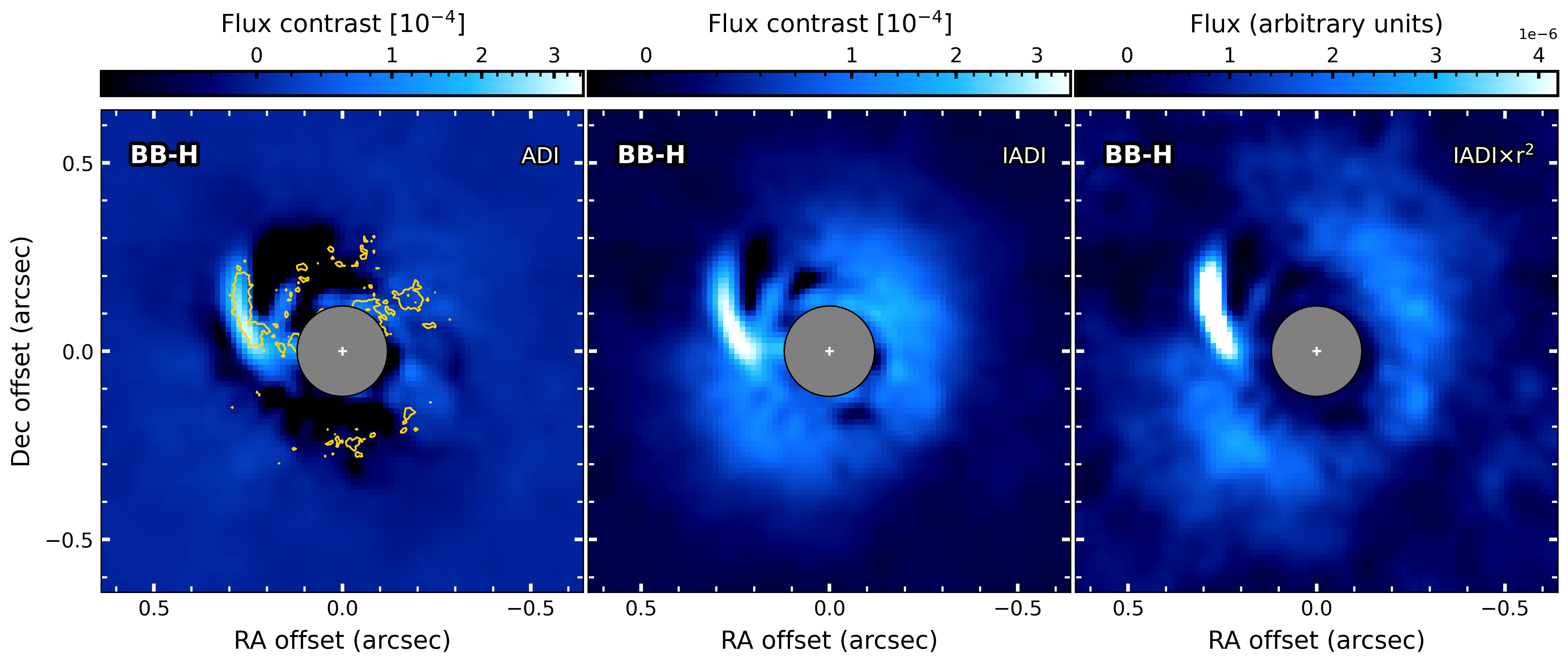}
    \caption{SHARK-NIR $H$-band images of CQ Tau (North on top, East on the left, square root color scale). \textit{Left:} PCA-ADI with 2 components. The \sv\ ADI \ha\ contours at 8$\sigma$ from Fig. \ref{fig:adi_iadi_r2} are overlaid for comparison. \textit{Middle:} PCA-IADI with 2 components and 70 iterations. \textit{Right:} PCA-IADI multiplied by $r^2$. The central darkened circle marks the area covered by the coronagraph.}
    \label{fig:nir}
\end{figure*}

As for the SHARK-VIS data, we applied both PCA-ADI, with 2 components, and PCA-IADI, with 2 components iterated 70 times (Fig. \ref{fig:nir}).

\section{Results}
\label{sec:results}

\subsection{SHARK-VIS and SHARK-NIR images} 
\label{sec:images}

The \sv\ \ha\ and continuum images are shown in Fig. \ref{fig:adi_iadi_r2}, where the central 200$\times$200 pixel area (1\farcs3 $\times$ 1\farcs3) is displayed.
The rows from top to bottom display the PCA-ADI, PCA-IADI, and PCA-IADI multiplied by the squared distance from the star $r^2$, projected onto the inclined outer disk plane, assuming an inclination $i=35$\degr and a position angle of $PA=55$\degr \citep{curone25}. From now on we will refer to these images simply as ADI, IADI, and IADI$\times r^2$. The $r^2$ multiplication is performed to compensate for the decay of disk illumination with distance, so as to enhance the structures seen in scattered light in the outer regions and reduce the extreme signal dynamics at smaller radii, allowing us to appreciate the whole disk in a single image.
It is worth noting that the central $45~\rm mas$ dark region in the IADI$\times r^2$ is not physical, but an artifact due to the PCA attenuation close to the star, which is not compensated by the $r^2$ illumination correction. For this reason, this innermost region contains unreliable information and we masked it in Fig. \ref{fig:adi_iadi_r2}.
The emission within this central region can be recovered with an optimized ad-hoc procedure that will be discussed in Sect.~\ref{sec:binary}. 
We also note that in Fig. \ref{fig:adi_iadi_r2} different features are better recognizable in one of the three rows than in the others, due to the high brightness range which makes it impossible to simultaneously render all the features with the same visual contrast.

The \ha\ ADI image presents bright emission from several regions, in particular from the prominent NE spiral arm and from additional spiral sections and arcs in the northern side of the target. In the \ha\ ADI image, we traced contours at 8 $\sigma$ with respect to the background standard deviation. The same \ha\ contours are overlaid on the continuum result, showing that the two images are basically consistent. Indeed, although the continuum image is noisier than the \ha\ image, the brightest emission regions observed in \ha\ are also detected in the continuum.

The IADI images recover the same emission present in the ADI images showing more continuous structures. In addition, they reveal further arms/arcs in the outer NW region. This is even more apparent in the IADI$\times r^2$.

The IADI results show again that all the features present in \ha\ are also observed in the continuum, thus confirming that in the \ha\ image we are observing scattered continuum light from extended disk structures. 
We conclude that we have no evidence of \ha\ excess emission from point-like sources and rule out accreting companions down to our sensitivity limit (Sect.\ref{sec:ha_contrast}). More generally, the \sv\ images do not show any point-like source outside of the $45~\rm mas$ central area. 

The SHARK-NIR $H$-band image (1\farcs3 $\times$ 1\farcs3 central area) is shown in Fig. \ref{fig:nir}, where we report again the ADI, IADI, and IADI$\times r^2$ processing results.
The contours from the \sv\ \ha\ ADI image are traced over the SHARK-NIR ADI image, showing that the two instruments provide consistent results for the bright NE spiral arm and the other spiral sections north-east of the star. The SHARK-NIR ADI already displays rather continuous structures, so that the application of the IADI, while recovering some scattered diffuse emission and mitigating the self-subtraction effects, does not seem to reduce the speckle noise as in the case of the \sv\ images. This difference is probably related to the different number of frames of the two datasets.

A detailed description of the disk structures detected in the SHARK images and a comparison with 
earlier high-contrast datasets \citep{uyama20,hammond22} is presented in Sect.\ref{sec:disk}.

\subsection{\ha\ contrast curve and limits on \ha\ emission}
\label{sec:ha_contrast}

\begin{figure}[t!]
    \centering
    \includegraphics[width=1\linewidth]{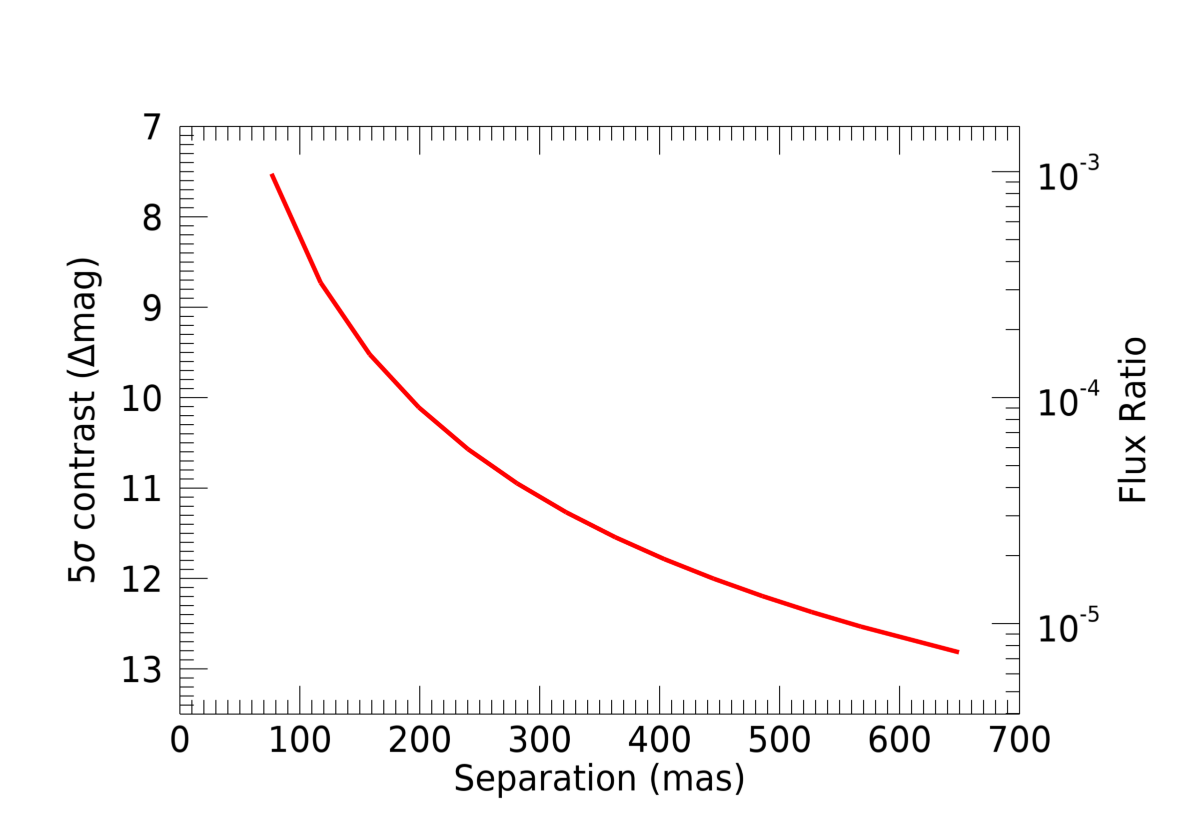}\\
    \includegraphics[width=1\linewidth]{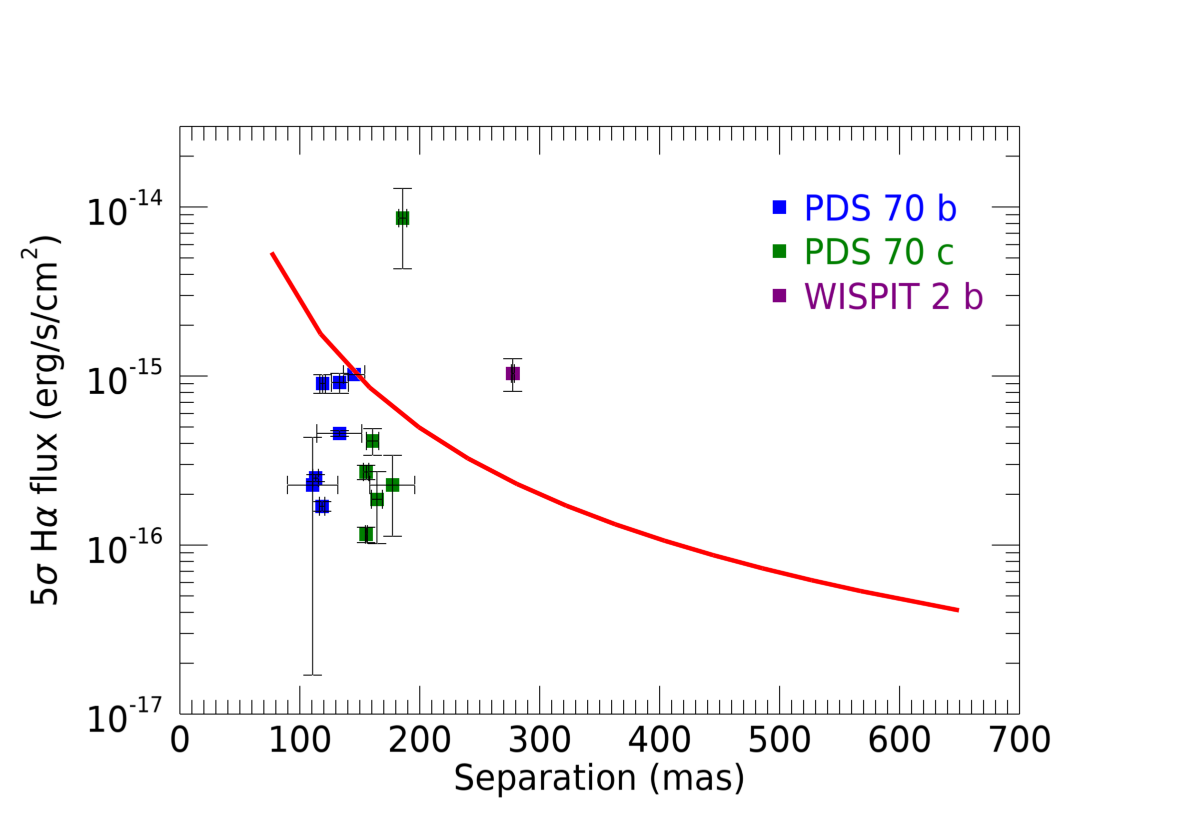}

    \caption{\textit{Top:} 5$\sigma$ contrast curve of the \sv\ \ha\ observation.
    \textit{Bottom:} \ha\ upper flux limit curve; available \ha\ measurements with their errors for known accreting protoplanets, rescaled in flux and separation to match the larger distance of CQ Tau, are overlaid for comparison (see text for references).
    }
    
    \label{fig:cc}
\end{figure}

We evaluated the $5\sigma$ contrast limit of our H$\alpha$ observations following a fake-planet injection procedure based on the recipe of \citet{mawet14}.
First, the IADI processed image was divided into a set of circular annuli with radius $r_i$. We then injected a dense pattern of synthetic planets in the non-derotated frames at different separations, contrasts, and different position angles. The number of planets per annulus was chosen to mutually avoid superposition through FOV rotation and to have all planets at least 2-FWHM away from each other. For each injected planet, $j$, the contrast $C_j$ is proportional to the radial profile $I(r)$ of the non de-rotated average stack, $C_j(r_i)=\alpha_j I(r_i)$, in order to have brighter fake companions at closer separations. We repeated this injection procedure five times, changing the proportionality coefficients $\alpha$ from $10^{-1}$ to $10^{-3}$ to generate a grid of contrast curves to interpolate. For each set, we ran the same IADI process used in data reduction. We then computed the average S/N of the planets for each annulus and alpha, S/N$(C_j, r_i)$, as the average of the S/N of their individual photometries. Finally, the S/N$(C, r)$ was interpolated to an S/N equal to 5, thus providing the $5\sigma$ contrast curve $C(r)$ of our H$\alpha$ (1 nm) observations, which is reported in Fig.~\ref{fig:cc}a.

To express the sensitivity of our observations in terms of \ha\ flux, we converted the contrast curve into an \ha\ flux upper-limit curve, which is shown in Fig. \ref{fig:cc}b. As a flux reference, we used a calibrated high-resolution spectrum of CQ Tau obtained in November 2017 with the HARPS-N instrument at the Telescopio Nazionale Galileo (TNG) in the framework of the GHOsT project \citep{gangi20,alcala21}.

Since the reference spectrum was not acquired simultaneously with our \sv\ observations, we searched for available information on the photometric state of the target at the time of our observations. CQ Tau is indeed an UX Ori-type variable star exhibiting frequent, stochastic photometric fluctuations, which are attributed to transient obscuration of the star by clumps of dust in the inner (likely warped) disk, which transit along the line of sight \citep[e.g.][]{grinin23}. 
We consulted the AAVSO database\footnote{American Association of Variable Star Observers International Database, https://www.aavso.org.} for all available $V$ and $R$-band CQ Tau photometry. For the date of our observation (Feb 17 2024) we found two $V$-band photometric points at 1:25 and 1:26 UT (our data were taken approximately between 2:00 and 4:19 UT) reporting $V$ = 9.531 and 9.558 mag with a 0.005 mag uncertainty, respectively. Another available $V$-band photometric point on Feb 17 21:39 UT is 9.514 mag (0.001 mag uncertainty), indicating that there was no fast trend in the magnitude variation. Assuming a $V$ magnitude of 9.55 at the moment of the \sv\ observations, we compared this value with the one used for the calibration of the reference GHOsT spectrum, which was $V$ = 9.84 mag \citep{gangi20}. This indicates a difference of 0.29 mag or a factor 1.31 in flux in the V band.
Past photometry of CQ Tau \citep[e.g.][]{berdyugin90,shakhovskoj05} as well as the AAVSO contemporaneous $V$ and $R$ magnitudes collected during the latest years show that the $V-R$ color is constant (0.45 mag) in this range of magnitudes, so we assumed that the same factor applies for the continuum around the \ha\, wavelength and used it to rescale the reference spectrum.

The conversion of the contrast curve was performed by integrating the reference spectrum over the wavelength range of the \sv\ 1 nm H$\alpha$ filter, and then multiplying the resulting flux by the contrast curve, which represents the ratio between the stellar flux and the flux of a detectable companion. This yields the upper-limit curve for the \ha\ emission from a possible protoplanet, that is reported in the bottom panel of Fig. \ref{fig:cc}.

On the same plot, we overlaid all actual \ha\ flux measurements available in the literature from the known accreting protoplanets (Fig. \ref{fig:cc}, bottom panel). These include \ha\ fluxes from PDS 70 b and c \citep[][and references therein]{wagner18, zhou25, close25a}, which indicate variable \ha\ emission from these planets, and the recent observation of WISPIT 2 b \citep{close25b}. The fluxes and separations have been rescaled to match the (larger) distance of CQ Tau, to provide a more meaningful physical comparison. 
The plot shows that we would have been able to detect \ha\ from a WISPIT 2 b-like planet and detect objects like PDS 70 b and c only in the brightest states of their variable \ha\ emission.

Finally, the \ha\ upper-limit curve can be converted into upper limits for the accretion luminosity $L_{\mathrm{acc}}$, 
by adopting the relationships between \ha\ luminosity and $L_{\mathrm{acc}}$ derived for accreting planets by \citet{aoyama21}. This provides $L_{\mathrm{acc}}$ upper limits of the order $10^{-5}$ \lsun\ at 350 mas and $10^{-4}$ \lsun\ at 100 mas. For this conversion, we assumed a visual extinction $A_\mathrm{V}$=0.5, as measured on the star \citep{gangi22}. However, it is possible that the extinction is higher in the local environment of a possible planet, in which case the corresponding accretion luminosity values would be larger.

\subsection{SHARK-NIR contrast curves and mass limits}
\label{sec:nir_contrast}

\begin{figure}[t!]
    \centering
    \includegraphics[width=0.88\linewidth]{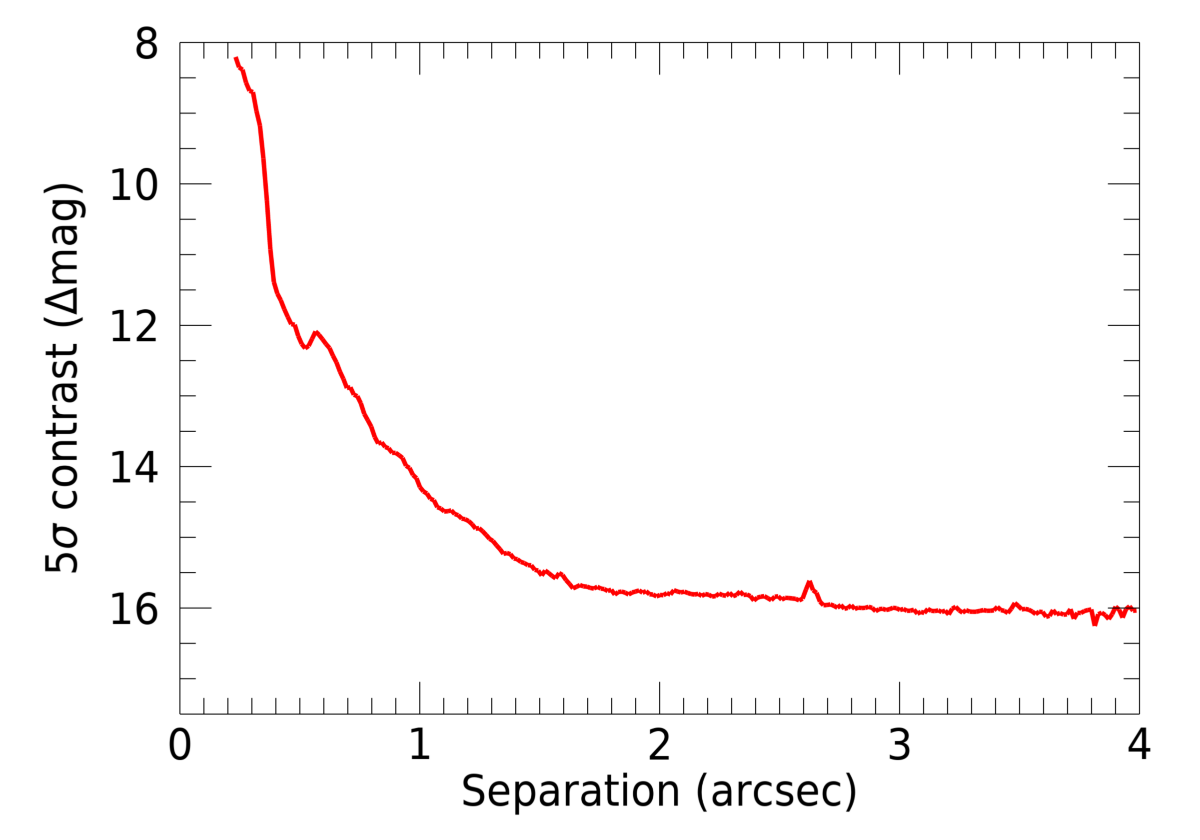}\\
    \includegraphics[width=0.88\linewidth]{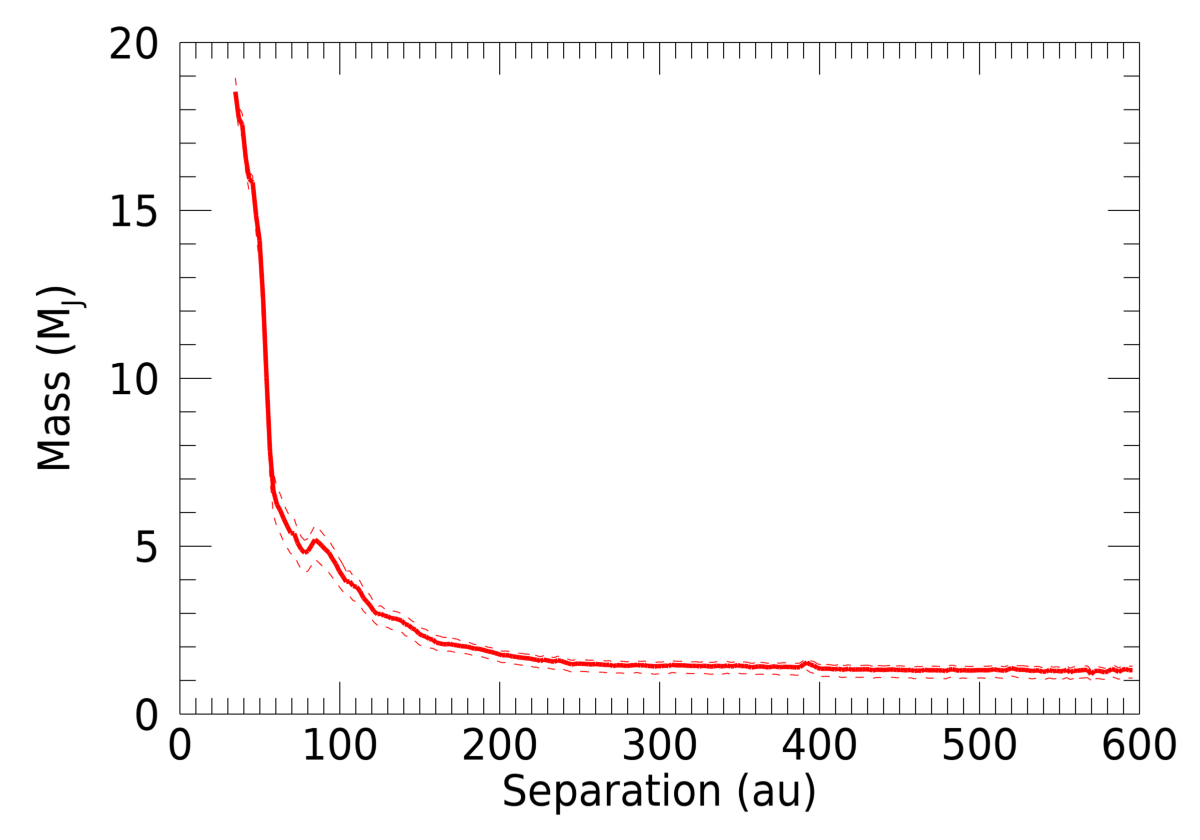}
    
    \caption{{\it Top:} 5$\sigma$ contrast curve for the SHARK-NIR broad-band $H$ data.
    {\it Bottom:} Mass limits derived from the contrast curve (see text for details) assuming an age of 10 Myr (solid line) and 8 and 12 Myr (dashed lines).
    }
    \label{fig:contnir}
\end{figure}

The contrast limit curve for the SHARK-NIR observation was calculated by estimating the
standard deviation within $1.5 \lambda / D$-wide annuli centered on the star. The self-subtraction caused by the post-processing method was evaluated by including simulated objects of known brightness at different separations from the host star.
The contrast was then consequently corrected. Finally, the results were corrected taking into account the effect of the small sample statistics as defined
by \citep{mawet14}. The final results of this procedure are displayed on the top panel of Fig.~\ref{fig:contnir}, where we can see that we achieve contrasts better than 14~mag at separations of the order of $\sim1"$ while we obtain contrasts of the order of
16~mag at separations larger than $1\farcs5$. \par
Starting from the derived contrast we evaluated the mass limits around CQ\,Tau using the AMES-COND models \citep{2003IAUS..211..325A} and assuming an age of 10$\pm$2 Myr and a distance of 149.4 pc (Sect.~\ref{sec:intro}). The results of such procedure are displayed in the plot in the bottom panel of Fig.~\ref{fig:contnir} where one can see that we are able to exclude companion with a mass larger than 5~\mj\ at
separations larger than $\sim70$~au from the star. Our limits are consistent with the results from \citet{hammond22}, where they report upper limits of $11~M_{\rm J}$ and $4~M_{\rm J}$ outward of $60~\rm au$ in the $L'$ and $Ks$ bands, respectively.
At smaller separations we are able to exclude companions with masses larger than 
$\sim19$~\mj, improving on the values obtained from the SPHERE data \citep{hammond22}.

\subsection{Innermost region: a close-in binary companion candidate}
\label{sec:binary}

\begin{figure*}[t!]
    \centering
    \includegraphics[width=0.8\linewidth]{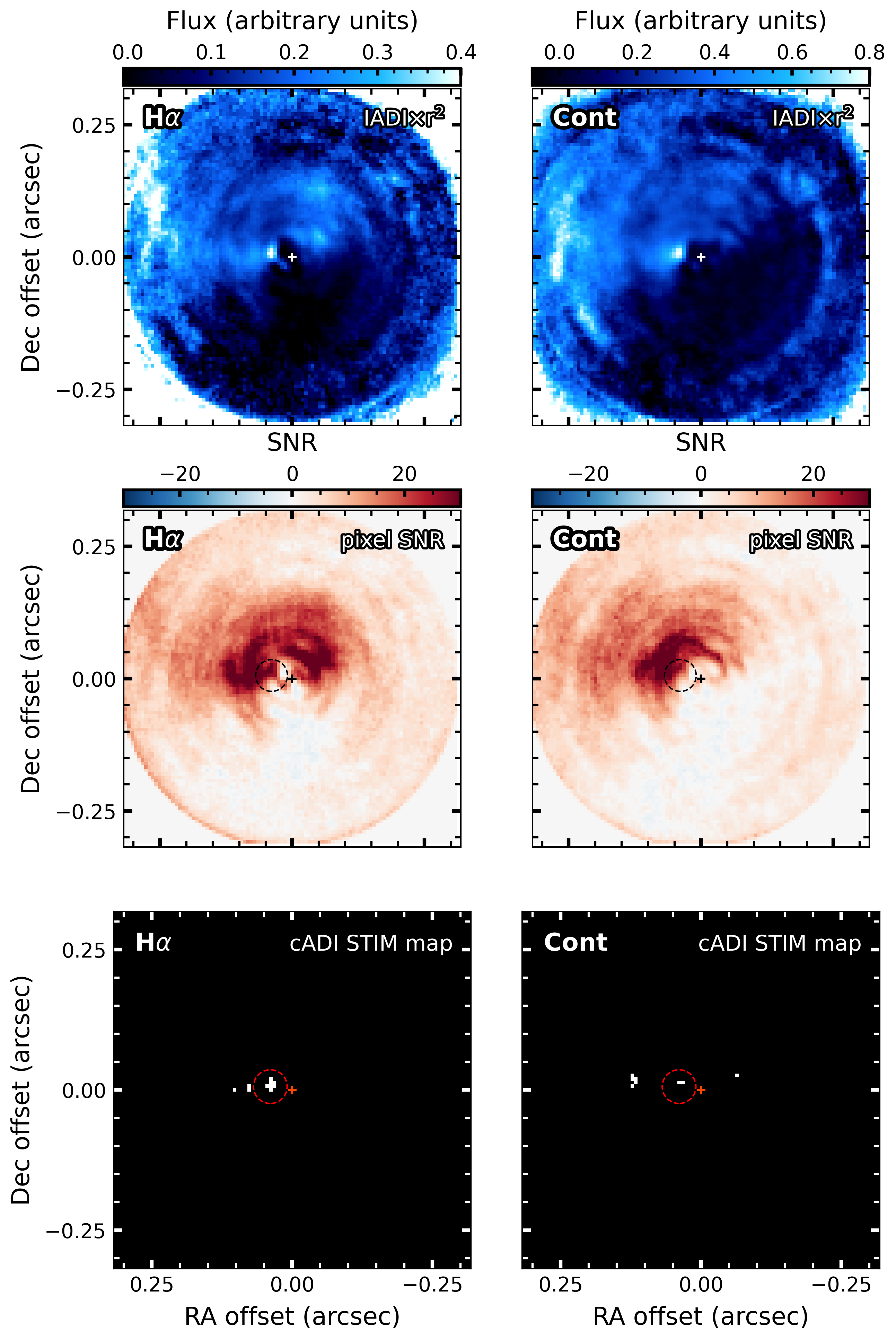}
    \caption{\textit{Top:} SHARK-VIS \ha\ (\textit{left}) and continuum (\textit{right}) IADI$\times r^2$ from the optimized processing for the innermost region around the star, showing evidence of the presence of a point-like companion. Note that the asymmetric PSF of the companion is a consequence of the multiplication by $r^2$.
    \textit{Middle:} Pixel SNR map for the \ha\ (\textit{left}) and continuum (\textit{right}) images, using a symmetric color scale ranging from -30 to 30 (red for positive values and blue for negative values). 
    \textit{Bottom:} Normalized STIM maps for the \ha\ (\textit{left}) and continuum (\textit{right}) classical ADI results (see text for details), displayed as binary black-white images (white color denotes signal surpassing the threshold of 1). In both the middle and bottom panels the location of the companion is marked using a dashed black and red circle, respectively, while the cross marks the center of the image.
    }
    \label{fig:binary}
\end{figure*}

SHARK-VIS ADI and IADI images shown in Fig. \ref{fig:adi_iadi_r2} were processed using almost all frames to maximize the signal, as described in Sect. \ref{sec:visobs}, and a high number of PCA components to optimize the detection of planets and structures at large separations. However, this procedure produces a strong attenuation of the signal at the image center, in a region with a radius of $\sim$ 45 mas, as already mentioned while describing Fig. \ref{fig:adi_iadi_r2}.

For this reason, we performed a separate processing specifically tailored for the innermost region around the star.
To do so, we maximized image sharpness rather than throughput by selecting only the few percent frames with the highest image quality, namely with the lowest FWHM, lowest ellipticity, and highest peak intensity. 
This resulted in a final dataset of 687 \ha\ frames and 94 continuum frames, which are however distributed over the whole range of field rotation angles, for a total integration time of 343 s and 47 s, respectively. The average FWHM of this subset is 24 mas (3.7 pixels), corresponding to a significantly higher Strehl ratio of 20\%.
We then cropped the central 100 × 100 pixels of the frames and performed the same ADI analysis using 2 PCA components only. We iterated this ADI for 50 times to get an IADI result, which we multiplied by $r^{2}$ thus revealing the central region, leaving only a small radius of $\sim 15$ mas that is still highly attenuated (see Fig.~\ref{fig:binary}, top panels). 
These images clearly reveal a bright point-like source located at the same position in both \ha\ and continuum images, which we interpret as a candidate companion star (see Sect. \ref{sec:mass}). 

To ascertain that this compact signal is not an artifact produced by speckle noise, we performed a number of checks and statistical analyses.
As a first check, we repeated the IADI analysis of the central region using the original and less aggressive frame selection used in Sect.\ref{sec:visobs}. The companion is still retrieved in both filters and the quantitative results that we derived from the two different selections are compatible within the error bars. Having confirmed that the results are independent of the selection criteria, as the stringent selection provides a slightly sharper PSF, we have used the images obtained with this selection for the subsequent analysis and checks.

The candidate is the brightest source in the inner part of the image after subtracting the central star, both for IADI and IADI$\times r^2$. 
We estimated the signal-to-noise ratio (SNR) of its signal in the IADI image as the aperture photometry (with a 1.5-pixel aperture radius) centered on the companion divided by the standard deviation of analogous aperture photometry measurements across a regular 9$\times$9 grid of locations in the image (centered on the companion and spaced by 7 pixels). Using this procedure, we obtain SNR values of 16.5 and 22.2 for \ha\ and continuum, respectively.

We also used a different approach by producing a map of the SNR for each pixel by dividing the pixel value in the IADI image by its standard deviation in the frame stack, whose average combination produces the IADI itself. The pixel SNR maps are shown in the middle panels of Fig.~\ref{fig:binary}. The SNR values we obtain for the pixel at the peak of the candidate PSF are 26.1 and 19.5 for \ha\ and continuum, respectively. 
The maps also show that the positive signal at the candidate location is unlikely to be a statistical fluctuation, as we do not find analogous negative peaks.

In addition, to estimate the statistical significance of the point-like source as a candidate detection, we applied the normalized Standardized Trajectory Intensity Mean (STIM) metric, as described by \citet{pairet19}.
To be conservative, we computed the STIM map for a classical ADI (cADI) image \citep{marois2006}, to evaluate the significance of the candidate detection even at the most basic level of processing, avoiding any possible attenuation induced by the PCA and obtaining a noise profile unaffected by iterative processing. 
We stress that the so-called inverse STIM map used for normalization (which we computed from the residuals using the opposite rotation angles compared to the actual observation) in our case definitely overestimates the speckle noise, as our frames are contaminated by the disk emission. Even assuming this larger speckle noise, however, the normalized STIM maps for both the \ha\ and continuum cADI images (reported in the bottom panels of Fig.~\ref{fig:binary}) show that the signal at the location of the candidate is significant \citep[normalized STIM $>$ 1, see][]{pairet19} in both filters.

Finally, to provide further support to our conclusions, we ran an additional test in which, for each filter, we subdivided the IADI$\times r^2$ frames in four subsets with the same number of frames and stacked them separately. We found that the candidate companion is consistently detected at the same location and contrast in all the subsets of both filters, even though at different sharpness depending on the varying seeing conditions.

After checking that the cadidate signal is not an artifact due to noise fluctuation,
we employed the Negative Fake Companion \citep[NFC,][]{Lagrange2010} technique to measure the position and contrast of the companion.
Since the NFC result might be affected by the distribution of the unknown background emission (for instance, due to disk features) at the candidate location, we decided to run the NFC procedure on the ADI, IADI, and IADI$\times r^2$ images for both \ha\ and the continuum. We eventually assumed the average and dispersion of the values thus obtained as the contrast and its uncertainty, respectively. Results are reported in Table \ref{tab:binary params}.

\begin{table}[t!]
\begin{center}    
\caption{Position and contrast of the candidate binary companion.}
\begin{tabular}{lccc}
\hline
\hline
Filter    &     Separation       &     Position Angle    &   Observed Contrast$^a$  \\ 
          &        [mas]         &        [deg]          &     ($\times10^{-2}$)  \\ \hline 
\ha\      &     $27.5 \pm 1.5$    &     $75.4 \pm 0.6$    &   $2.0 \pm 1.0$   \\
Cont      &     $27.3 \pm 1.5$    &     $75.2 \pm 1.0$    &   $4.0 \pm 1.0$ \\ \hline
\end{tabular}
\\
\end{center}
\tablecomments{~$^a$ directly measured contrast; this is affected by obscuration of the primary (see Sect.\ref{sec:mass} for details).}
\label{tab:binary params}
\end{table}

The measured contrast with respect to the central star is $(2.0 \pm 1.0) \cdot 10^{-2}$ in \ha\ and $(4.0 \pm 1.0) \cdot 10^{-2}$ in continuum. The different contrast results in the two filters indicate that the \ha\ emission from the companion is weaker than that from the primary \citep[see e.g.][]{alcala21,gangi22}.
The position of the point-like emission from the candidate is the same within errors in both filters, which is an additional indication that this is not an artifact.
The companion average separation is $27.4 \pm 1.5$ mas, with an average position angle of $75.3 \pm 0.8^\circ$. 

Assuming a distance of 149.4 pc \citep{gaiadr3} for CQ Tau, the measured separation corresponds to a distance of 4.1$\pm0.2$ au on the plane of the sky and of 5.0$\pm$0.3 au on the plane of the outer disk \citep[$i = 35\degr$;][]{curone25}.

On the basis of our single measurement, we cannot completely rule out that the source is an object that is not physically bound to CQ Tau. However, we remark that a similar star would have been easily spotted in previous direct-imaging campaigns, when it would have been at larger separations due to the CQ Tau proper motion \citep[2.987 and -26.364 mas/yr,][]{gaiadr3}. 

\vspace{1cm}
\subsection{Mass of the candidate companion}
\label{sec:mass}

To derive an estimate of the companion mass we consider its thermal emission contrast with respect to the primary. As the \ha\ emission from the companion and primary could bias this measurement, we rely on the contrast observed in our 10 nm continuum filter centered at 647 nm.

An important point to consider is that CQ Tau, as mentioned in Sect.~\ref{sec:ha_contrast}, is a well known UX Ori variable subject to photometric variations due to clumps of dust transiting along the line of sight. 
The compact misaligned and possibly warped inner disk observed by \citet{bohn22} through near-IR interferometry (semimajor axis $a=1.16$ mas, corresponding to 0.173 au) is the likely source of these obscuration effects. Our candidate companion is clearly located outside of this inner disk, so the obscuration should only affect the primary. 
Consequently, the observed contrast in the continuum requires a correction as we are not observing the primary at its maximum brightness, which represents the emission from the unobstructed photosphere. 

From archival photometry \citep{berdyugin90,shakhovskoj05,grinin23} and the AAVSO data already mentioned (Sect. \ref{sec:ha_contrast}), we derive that the maximum $V$ and $R$ magnitudes for CQ Tau are $\sim$ 8.70 and 8.35 mag, respectively. At the moment of our observation we had $V$=9.55 mag, corresponding to $R$=9.1 assuming the $V-R = 0.45$ mag indicated by the same photometric datasets. This leads us to estimate a 0.75 mag obscuration in the $R$ band at the time of our observation, which translates into a factor of two in flux variation. Applying this correction to our measurement  
we obtain that the continuum flux ratio between the unobscured primary and the companion is 0.020$\pm$0.005.

To estimate the binary mass ratio from this contrast, we considered blackbody emission for both the primary and the secondary stars, characterized by their effective temperatures and radii. 
The range of parameters adopted for the primary star were as follows: $T_{\mathrm{eff}} = 6823 \pm 136$ K \citep{alcala21}, $R_\star = 2.0 \pm 0.2$ $R_{\odot}$ \citep[e.g.][]{mendigutia12,vioque18,ubeira-gabellini19,gangi22}, and age = $10 \pm 2$ Myr \citep[e.g.][]{mendigutia12,ubeira-gabellini19}.

For the secondary component, we considered the \citet{baraffe15} pre-main-sequence models at the assumed age of the primary, which provide $T_{\mathrm{eff}}$ and stellar radius as a function of the stellar mass. 
This allowed us to derive the expected flux ratio between the emission of the primary and the secondary in our filter at 647 nm for all combinations of the parameters.
By comparing these predicted values with our contrast measurement, corrected for the primary obscuration, we obtain a final range of possible masses for the candidate companion.
Accounting for the associated uncertainties,
we eventually derive a mass range 0.46–0.85 $M_{\odot}$.

We note that in the unlikely scenario in which the companion is also affected by a residual extinction from the inner disk, we would obtain a higher upper mass limit, which would reach 1.05 \msun\ in the limit where both stars experience the same obscuration. However, this higher mass value seems highly unlikely given also the constraints from Gaia (Sect.~\ref{sec:discussion:companion}).

Considering a mass range 0.46–0.85 \msun\ for the companion and a mass of the primary between 1.5-1.7 \msun (Sect.~\ref{sec:intro}) we can calculate the resulting semi-major axis of the binary orbit using Kepler's third law by assuming an orbital period. Recent analysis of the historical light curve of CQ Tau by \citet{grinin23} \citep[see also][]{shakhovskoj05} has shown the presence of a clear 10 yr periodicity, which the authors link to the period of a companion. Assuming a 10 yr period, with the masses mentioned above we obtain semi-major axes spanning in the 5.8-6.3 au range.

This estimate represents a preliminary assessment pending follow-up observations which will definitely be required to confirm the nature of the candidate.
In Sect.\ref{sec:discussion:companion}, we will further discuss the candidate binary in the context of recent literature results.

\subsection{Disk structures}
\label{sec:disk}

\begin{figure*}[t!]
    \centering
    \includegraphics[width=0.92\linewidth]{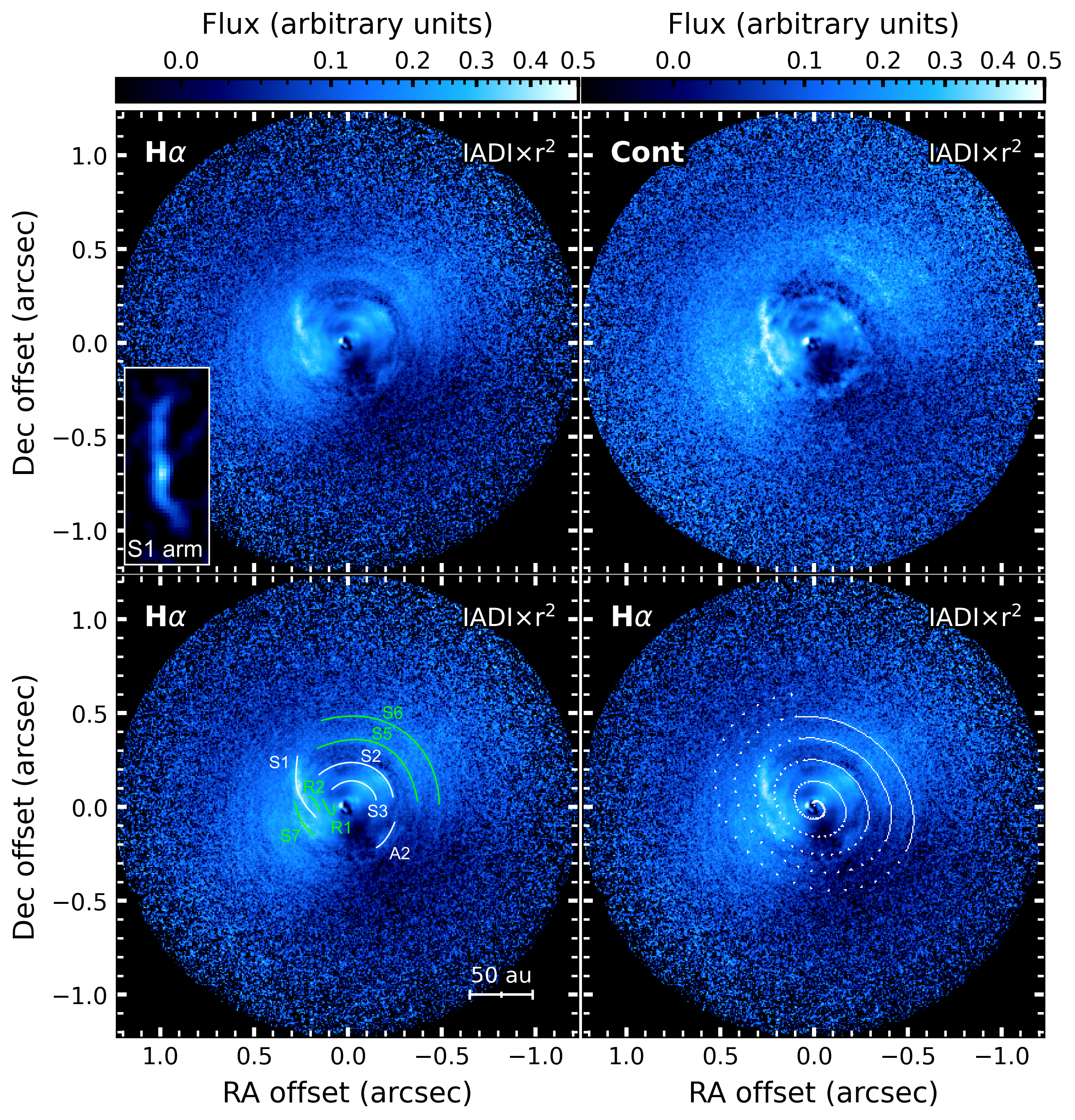}
    
    \caption{
    \textit{Top left:} synoptic view of the CQ Tau system obtained by compositing the \sv\ \ha\ IADI$\times r^2$ image optimized for the innermost region (from Fig. \ref{fig:binary}) with the one optimized for the external regions (from Fig. \ref{fig:adi_iadi_r2}); the inset shows a zoom of the NE arm (S1), after application of a 15-pix high-pass filter and 2-pix Gaussian smoothing, which clearly shows the wiggling of the NE arm described in the text.
    \textit{Top right:} the same composite image for the continuum filter. 
    \textit{Bottom left:} main structures observed in the CQ Tau disk, marked over the \ha\ composite image of the first panel.
    \textit{Bottom right:} the trace of a single Archimedean spiral projected onto the inclined flared disk surface is marked over the same image.
    \label{fig:synoptic_CQTau}
    }
\end{figure*}

In order to produce a visually enhanced and comprehensive view of CQ Tau disk, we blended the \sv\ images optimized for the central region (Fig. \ref{fig:binary}) into the central $45~\rm mas$ radius area of the full-field IADI$\times r^2$ images (Fig. \ref{fig:adi_iadi_r2}). We did this for both filters after matching the average flux of the inner and outer images over an annulus extending from $45$ to $100$ mas. This provided the two composite images (\ha\ and continuum) reported in Fig.~\ref{fig:synoptic_CQTau} (top row), which show an unattenuated synoptic view of the CQ Tau disk from a radius $\sim$ 780 mas down to the very center where the candidate binary companion is found.

We used the \ha\ composite image to mark and label the main disk structures retrieved by the SHARKs (bottom left panel of Fig. \ref{fig:synoptic_CQTau}). 
The three northern spiral arms, which we labeled S1, S2, and S3 with reference to the paper by \citet{hammond22} based on infrared data from the Very Large Telescope (VLT), are clearly recovered.
The S1 arm is the most prominent feature not only in the infrared, but also in the visible. Its overall morphology and extent (about $0\farcs25$) globally follow those observed in the infrared \citep[see also][]{uyama20}. The small displacement observed between the $H$-band and $\sim$650 nm emission (see the contours in the left panel of Fig.~\ref{fig:nir}) presumably reflects the different height of the disk surface probed at the two different wavelengths.
The \sv\ images, thanks to their higher spatial resolution, reveal that the S1 arm is actually characterized by a few wiggles or twists (see inset in Fig.~\ref{fig:synoptic_CQTau}, top left), which are suggestive of locally complex dynamics.
In this regard, we note that \citet{hammond22} reported the presence of a twist in the S1 arm based on their NACO data, which they interpreted as an artifact due to the presence of bad pixels in the detector at the same location, or due to a geometric bias produced by the ADI processing. As we detect a very similar twist in the \ha\ and continuum data using both ADI and IADI, we instead propose that it is a real feature of the S1 arm and not an artifact.

Emission from the segments of spirals labeled as S2 and S3 is generally weaker but their overall shapes remain easily recognizable. Some flux variations are visible, which may correspond to localized regions of enhanced dust density. 
The S3 arm 
is located basically north of the source and its position is consistent with the inner northern arm identified by \citet{safonov22} using $I$-band differential speckle polarimetry.
We retrieve no evidence for the S4 arm reported by \citet{hammond22} SW of the star. Of the two arcs, A1 and A2, observed by \citet{hammond22} and located also SW of the source, we marginally detect only A2 in both visible and near-infrared, which appears to be a possible extension of the S2 arm.

A particularly bright emission region (labeled R1), which is evident especially in the ADI images,
is located E of the star at separations between $\sim 50$ - $130$ mas. There is evidence for additional emission bridging this region to the S1 and S2/S3 structures, which we label R2, observed in both the visible and near-infrared images; we cannot however determine if this reflects a true physical connection. This region has a shape that resembles a broken spiral arm.

The \sv\ IADI$\times r^2$ images reveal two additional faint arms/arcs at larger distance from the star, extending from N to W (S5 and S6), the latter also recovered in the SHARK-NIR IADI$\times r^2$ image, which visually appear as a possible extension of the S1 spiral arm.

Finally, we observe another arm section (S7) east of the star, which seems to be separated from the prominent S1 arm and was not previously reported.

Overall, our SHARK images show a disk morphology that is globally consistent with that derived from previous scattered-light observations, but reveal additional structures and morphological details.  
We discuss the nature of these structures and their possible mutual connections in Sect. \ref{sec:discussion:disk}. 
The location of the detected features also is indication that some sections of the disk (especially the southern half) receive less light from the star. We analyze more in detail these shadowed areas of the disk in Sect.\ref{sec:discussion:shadows} in the framework of the inner misaligned disks identified via interferometric observations \citep{bohn22}.

\section{Discussion} 
\label{sec:discussion}

\subsection{Nature of the disk structures} 
\label{sec:discussion:disk}

The multiple arms/arcs structures retrieved in our images can be interpreted as a series of spirals excited by a companion.
The formation of spiral features in circumbinary disks surrounding binaries is well documented in the literature, both from the theoretical/numerical \citep[e.g.][]{miranda17,ragusa17} and observational perspective \citep[e.g.][]{tobin16,alves19}.

From their analysis of infrared scattered-light images, \citet{hammond22} proposed that an unseen close-in companion on an inclined and eccentric orbit might explain the arms, the cavities, and the disturbed gas kinematics. Such a companion is indeed compatible with the candidate spotted by \sv.
In this scenario, the regions of enhanced emission might be directly connected to the binary companion either via density waves expanding in the disk or actual warps of the disk surface \citep[e.g.][]{ragusa24,winter25}. The distorted morphology highlighted for instance in the prominent S1 arm by the high angular resolution of the \sv\ images (see inset in Fig.~\ref{fig:synoptic_CQTau} top-left panel) is indicative of complex dynamics, possibly related to the gravitational disturbance of the massive companion.

Another possibility is that at least some of the observed spiral-like structures are not caused directly by the companion but are instead related to gas infalling from the ambient medium at later evolutionary phases, which can produce density waves and warping \citep[e.g.][]{winter25,calcino25}. 
Late infall has recently been proposed by \citet{garufi26} as the probable origin of spirals and warps observed in evolved massive disks like CQ Tau. 
Using HST/STIS, \citet{grady05} observed ambient material in the shape of an arc seemingly pointing at the source from large distances, which might be interpreted as a streamer. Interestingly, this putative streamer approaches the source from the N-NE direction, which is where \citet{Zagaria25} detected SO emission from the disk. This might be the signpost of an ongoing shock from material falling onto the disk \citep[e.g.][]{garufi22,speedie25}. The peak of this SO emission is located approximately in a position in between R1, S2, and S3 \citep[see Fig. 3 in][]{Zagaria25}. In this scenario, the scattered light coming from one of these structures might actually trace the end of the infalling streamer observed with HST.

A final possible interpretation is driven by the observation that there is a seemingly constant spacing between the S2, S3, S5, and S6 arms observed in the NW quadrant of the disk. This might suggest that these arms and possibly most of the other structures of the disk are actually part of a single grand-design spiral with constant pitch (Archimedean spiral).
We traced a possible spiral of this type over the disk surface \citep[assuming the flaring profile derived by][]{galloway-sprietsma25} and obtained the result shown in Fig. \ref{fig:synoptic_CQTau}. This spiral can indeed reproduce very well all the arms in the NW sector of the disk and the A2 arc, but it seems to trace less precisely the bright features in the eastern part of the disk (e.g. S1 and the initial section of the S2 arm), possibly because of the perturbations induced by the companion. 
Such a spiral feature is unlikely to originate from the excitation of density waves in the disk; were that to be the case, we would expect the constructive interference of different wave modes to produce a logarithmic spiral \citep[e.g.][]{ogilvie02,fasano24}.
Instead, Archimedean spirals have been associated for instance to mass loss in AGB binary outflows \citep[e.g.][]{maercker12}.
Despite the fundamental differences in stellar evolution stages between CQ Tau and AGB stars, we speculate that, if some material in CQ Tau is ejected at constant speed through a wind or Roche lobe overflow from the binary, the resulting geometry would be an evenly spaced Archimedean spiral. As this outflowing material climbs the flared, bowl-shaped disk surface, geometric projection distorts its apparent morphology into the observed structure.
It is unclear if a similar process, involving the candidate binary, might be at work in CQ Tau.

\subsection{Shadows} 
\label{sec:discussion:shadows}

\begin{figure}[t!]
    \centering
    \includegraphics[width=1\linewidth]{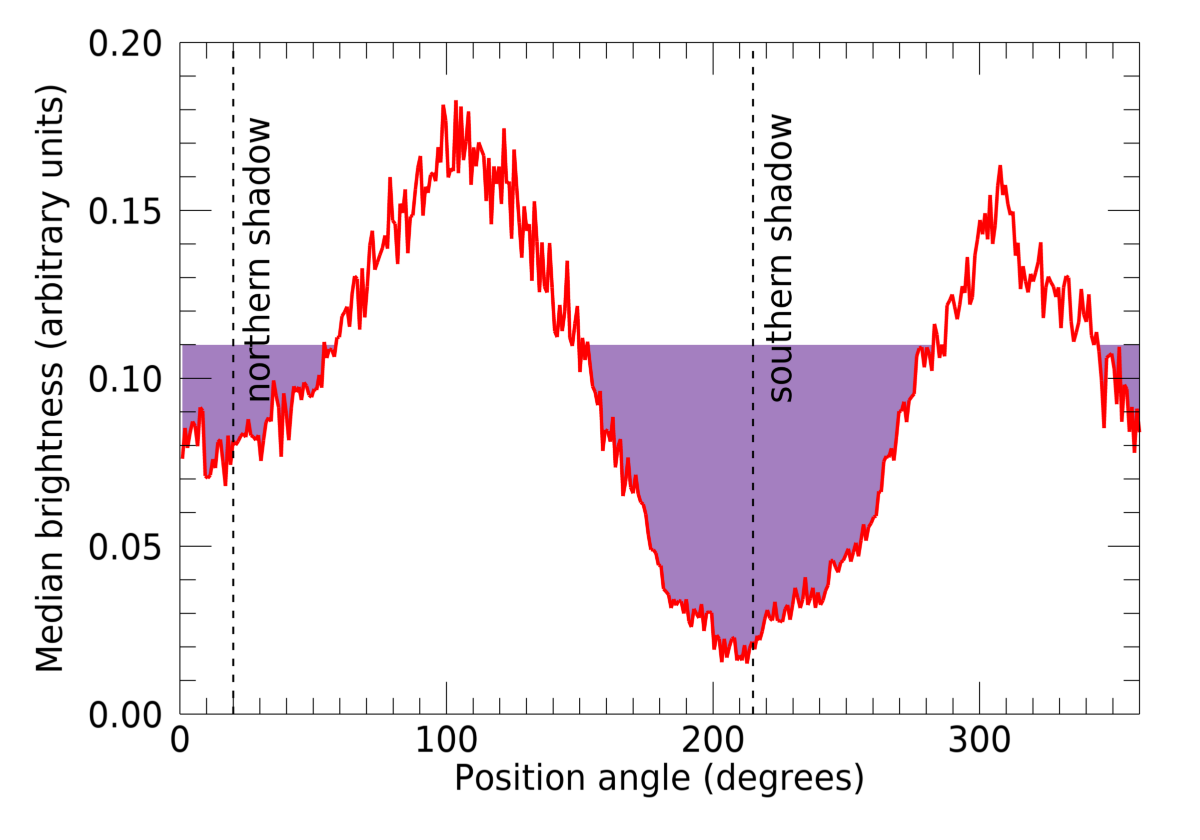} \\
    \includegraphics[width=1\linewidth]{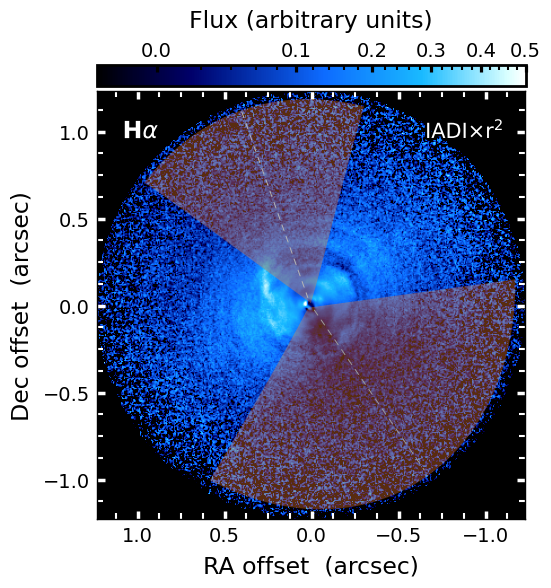}
    
    \caption{\textit{Top:} Median azimuthal brightness profile of the CQ Tau disk, showing the presence of two wide shadowed sectors (northern and southern shadows) with approximate axes at PA 20\degr and 215\degr.
    \textit{Bottom:} The shadowed sectors from the plot in the top panel are marked over the composite \ha\ image from Fig. \ref{fig:synoptic_CQTau}.}
    \label{fig:shadows}
\end{figure}

By inspecting the composite images in Fig. \ref{fig:synoptic_CQTau}, we can clearly see that the average disk brightness is noticeably reduced in a wide range of azimuth directions. We performed a median in the radial direction over the central area of the image, over a radial interval of 1\farcs2 from the central star, and obtained a brightness profile as a function of the position angle (Fig.~\ref{fig:shadows}, top panel). This displays two wide ($\sim$ 70\degr and 130\degr) shadows with axes approximately around PA $215$\degr\ and PA $20$\degr, with the southern shadow significantly deeper than the northern one; we schematically mark these shadowed sectors over the \ha\ composite image in the bottom panel of Fig.~\ref{fig:shadows}.
We infer that these sectors of the disk do not reflect light from the central star because they are mostly in the shadows.

Shadows are observed in a large number of disks and are typically associated with the presence of an inner misaligned disk
\citep[e.g.][]{min17,orihara25,bohn22}.
For CQ Tau, interferometric measurements obtained initially with PTI \citep{eisner04} and more recently with VLT/GRAVITY \citep{bohn22} provided evidence for a misaligned compact ($<$0.5 au) inner disk. 
Considering the positions of the shadows in our images, we find that these are in general agreement with the predictions of \citet{bohn22} for a misalignment angle $\Delta\theta=41$\degr ($\Delta\theta_2$ scenario, see their Table 4). Indeed the \citet{min17} framework used by the authors provides a distribution of the position angles of the south and north shadow axes peaked around 191\degr\ and 11\degr, respectively (see their Fig. 7, orange lines), which are relatively close to our measurements of 215\degr\ and 20\degr.
We note that the PA difference between the two shadows not being 180\degr\ is consistent with the geometric effect due to the shadows being projected over an inclined flared surface.

Wide azimuthal shadows can be connected to either a small misalignment between the disks \citep[e.g.][]{winter25,bohn22} (which, however, does not seem to be the case for CQ Tau based on the GRAVITY results) or to a large scale height and/or significant warp of the inner disk \citep[e.g.][]{orihara25}. A scenario with a highly asymmetric or perturbed inner disk  is compatible with the fact that the northern and southern shadows have a different depth and also with the observed UXor photometric behavior. 

Finally, the small change of the shadow axis angles between the VLT/SPHERE scattered-light images shown by \citet{bohn22} and our SHARK data might suggest an actual shift of the shadows in the time span between the two observations. To understand how and if this change is connected to the inner disk (for instance to its precession, to possible warps, or to its scale height) a complete and detailed simulation would be required, which is beyond the scope of this work.

\subsection{The candidate companion in the context of Gaia and ALMA observations} 
\label{sec:discussion:companion}

The presence of a binary companion as the one directly imaged by \sv\ is supported by indirect evidence from recent Gaia and ALMA data.

Gaia astrometric measurements of CQ Tau are characterized by a Renormalised Unit Weight Error (RUWE) that significantly differs from unity (RUWE = 3.7), which signals a strong astrometric deviation from a single-source solution \citep[e.g.][]{castro-ginard24}. Gaia data have been recently analyzed in detail by \citet{vioque26}, who investigated proper motion anomalies for a large sample of transitional disks. The authors derive the distribution of masses and semi-major axes of the possible companions responsible for these anomalies. For CQ Tau, they obtain indication for a massive companion with mass $309^{+840}_{-190}$ M$_\mathrm{J}$ and separation $0.99^{+4.8}_{-0.31}$ au (see their Table A.1 and Fig. A.2), which is formally compatible with our detection, though the properties of our candidate do not place it close to the peak of their distribution.

Using a statistical approach, \citet{ragusa25} examined a sample of transitional disks to study the likelihood that a companion carving the cavity has eluded detection thus far due to its small projected separation or faintness, finding that such a scenario is highly probable for CQ Tau.
Using their framework for CQ Tau, however, considering the binary with a 
separation of $\sim 4\mathrm{-}5$ au and the gas cavity of $\sim 20 \textrm{--} 25$ au, the fraction of companion configurations that would be able to carve the observed cavity is $\lesssim 20\%$. This low likelihood suggests that the observed candidate companion alone is unlikely to be responsible for carving the cavity, pointing toward the presence of a putative additional, unseen object at larger separations \citep[e.g.][]{ubeira-gabellini19}.

High-resolution observations of the gas kinematics with ALMA in the disk of CQ Tau have revealed deviations from Keplerian rotation and complex dynamics, with spirals \citep[e.g.][]{woelfer21,fukagawa26}, vertical flows in the gas \citep{benisty26}, velocity kinks \citep{Pinte25}. These features are consistent with the predicted signatures from a massive companion \citep[e.g.][]{calcino23, calcino24, ragusa24}, and have been interpreted also as the evidence for a moderate warping of the disk \citep{winter25}. 

One final interpretation of the bright features and shadow lanes that we observe across the disk surface (Fig. \ref{fig:synoptic_CQTau}, Sect. \ref{sec:disk} and \ref{sec:discussion:shadows}) is that these are indeed connected to warped areas of the disk. In this scenario, some regions might present a different angle to the stellar light and/or be at higher elevation with respect to the surrounding area, so that they intercept more starlight, or, on the contrary, remain in the shadows behind a warped rim. For instance, this might be the case of the A2 arc/arm, which is the only structure visible in the wide shadowed sector in the SW direction.

\section{Conclusions} 
\label{sec:conclusions}

In this paper we presented high-contrast observations of the pre-main sequence star CQ Tau and its transitional disk, taken at LBT with SHARK-VIS in narrow-band \ha\ and adjacent continuum at $647$ nm, and with SHARK-NIR in broad-band $H$, in the context of a survey program in Taurus-Auriga aimed at spotting and characterizing giant protoplanets in formation around young targets. 
The \sv\ data are the first high-contrast observations of this target in the visible band.

Our main results can be summarized as follows:

\begin{itemize}

    \item We detect no point-like sources in either \ha\ or $H$-band compatible with a planetary companion. From the \ha\ contrast curve, we derive an upper-limit flux curve for the \ha\ emission from possible accreting planets. The curve shows that we would have been able to detect \ha\ emission from a planet like WISPIT 2 b, while planets like PDS 70 b and c would have been detected only when their \ha\ variable emission is at maximum. From the near-infrared contrast curve, we derive limits on the mass of possible companions, ruling out objects with a mass larger than 5~\mj\ at separations larger than $\sim70$~au from the star, and larger than $\sim19$~\mj\ at smaller separations down to the region masked by the SHARK-NIR coronagraph.
    
    \item In the SHARK-VIS images, obtained without a coronagraph, we performed an optimized reduction to inspect the innermost region and identified a candidate close-in companion star at 27.4 mas separation, corresponding to 4.1 au (5.0 au deprojected on the plane of the outer disk) and PA 75.3\degr. From the observed contrast, we estimate a possible mass range of 0.46–0.85 $M_{\odot}$ for the companion. This candidate might be the massive companion invoked by previous works to explain the spirals and the disturbed kinematics observed in the disk. Its properties are also formally compatible with Gaia astrometric measurements. 
    Considering the small projected separation of the companion and the wide gas cavity of the disk (20--25 au radius), it is plausible that an additional unseen object at larger distance has also contributed to the carving of the cavity.

    \item Our images reveal a highly structured disk seen in scattered light. We confirm the global morphology already revealed by previous near-infrared high-contrast campaigns, recovering many spiral-like structures such as the prominent and bright NE arm and other arms in the northern section of the disk. 
    Our images unveil additional arms in the NW direction and, thanks to the higher resolution achieved with the visible-band data, show a series of twists in the morphology of the bright NE arm, indicating complex dynamics on the disk surface. These structures are possibly connected to the perturbation induced by the companion, or (at least in part) might arise from late mass infall from the ambient medium, a mechanism recently suggested for many evolved disks with spirals, such as CQ Tau. 
    An alternative interpretation is that all or most of the arms are part of a grand-design single Archimedean spiral (all the NW arms can be indeed traced with a similar spiral). However, to create such a structure, a mass loss ejection at constant speed (e.g. through a wind or outflow) must be invoked; it is unclear if such a mechanism can be at work in CQ Tau.
    
    \item We observe the presence of wide shadows over the disk. These are connected to the presence of the compact misaligned inner disk revealed by previous interferometric measurements, which might be the result of the gravitational effects of the orbiting companion. The position of the shadows is compatible with this inner disk, while their angular extent suggests that the inner disk might be rather thick and warped. This is also in agreement with the known UXor photometric variations of the source.
    We notice a slight variation of the shadows compared to previous near-infrared images (so on timescales of a few years), which however needs to be confirmed with new scattered light images.
    
\end{itemize}

Additional high-contrast observations of CQ Tau, possibly in multiple filters, are definitely needed to further investigate the scenarios we have outlined and in particular to validate the binary candidate seen by \sv, so as to derive a more accurate estimate of its mass and provide a first assessment of its orbit.
The upcoming Gaia DR4 release will provide multi-epoch measurements over a 5-year temporal baseline. A companion of $\sim$0.5 $M_{\odot}$ on an orbit with a semi-major axis of $\sim$6 au should produce an astrometric signature of the order of 10 mas, which is considerably larger than the single epoch accuracy of 15 $\mu$as expected for stars with G$<$13. DR4 might therefore indirectly confirm the presence of the companion and provide data to constrain its orbit. Validation through a second direct detection will be however fundamental. Assuming a period of about 10 yr and an orbit on the plane of the outer disk, we can expect a variation of the companion position of the order of 20-25 mas per year. On this basis, new observations in the visible with an angular resolution of $\sim$20 mas should already be able to detect a significant variation of the companion location with respect to early 2024.
Our work shows that high-contrast imaging in the visible is a powerful tool to reveal close-in companions that might remain undetected in typical near-infrared coronagraphic images.

\footnotesize
\section*{Acknowledgements}
This paper makes use of data obtained with the Large Binocular Telescope (LBT).
The LBT Corporation partners are: The University of Arizona on behalf of the Arizona university system; Istituto Nazionale di Astrofisica, Italy; The Ohio State University; The Research Corporation, on behalf of The University of Notre Dame, University of Minnesota and University of Virginia.

The SHARK-VIS and SHARK-NIR teams acknowledge with thanks the LBT crew and the LBT AO and telescope operators for their invaluable help and support during the observations.

We acknowledge the significant contribution of the
ALTA Center (alta.arcetri.inaf.it) forecasts, generated using the Astro-Meso-Nh
model. Initialization data of the ALTA automatic forecast system is sourced from
the General Circulation Model (HRES) of the European Centre for Medium-
Range Weather Forecasts.

We acknowledge the variable star observations from the American Association of Variable Star Observers International Database (https://www.aavso.org) contributed by observers worldwide and used in this research.

SA acknowledges financial contributions from PRIN-MUR "POPS" 2022YP5ACE and from the INAF GO grant "GO SHARK!".

ER acknowledges support from the European Union (ERC Starting Grant DiscEvol, project number 101039651). Views and opinions expressed are, however, those of the author(s) only and do not necessarily reflect those of the European Union or the European Research Council. Neither the European Union nor the granting authority can be held responsible for them.

\bibliography{cqtau}{}
\bibliographystyle{aasjournal}



\end{document}